\documentclass[iop]{emulateapj}
\usepackage{natbib}

\usepackage{amsmath}
\usepackage{graphicx}
\usepackage{url}

\providecommand{\tab}[1]{Table~\ref{tab:#1}}
\providecommand{\fig}[1]{Figure~\ref{fig:#1}}
\providecommand{\refsec}[1]{Section~\ref{sec:#1}}

\begin{document}
\title{Improved Spectrophotometric Calibration of the SDSS-III BOSS Quasar Sample}

\author{Daniel Margala\altaffilmark{1}, David Kirkby\altaffilmark{1}, Kyle Dawson\altaffilmark{2}, Stephen Bailey\altaffilmark{3}, 
Michael Blanton\altaffilmark{4}, Donald P. Schneider\altaffilmark{5,6}}
\altaffiltext{1}{Frederick Reines Hall, Department of Physics and Astronomy, University of California, Irvine, CA, U.S.A.}
\altaffiltext{2}{Department of Physics and Astronomy, University of Utah, Salt Lake City, UT 84112, USA}
\altaffiltext{3}{Lawrence Berkeley National Laboratory, One Cyclotron Road, Berkeley, CA 94720, USA}
\altaffiltext{4}{Center for Cosmology and Particle Physics, Department of Physics, New York University, 4 Washington Place, New York, NY 10003, USA}
\altaffiltext{5}{Department of Astronomy and Astrophysics, The Pennsylvania State University, University Park, PA 16802}
\altaffiltext{6}{Institute for Gravitation and the Cosmos, The Pennsylvania State University, University Park, PA 16802}

\email{dmargala@uci.edu}

\begin{abstract}
We present a model for spectrophotometric calibration errors in observations of quasars from the third generation of the Sloan Digital Sky Survey (SDSS-III) Baryon Oscillation Spectroscopic Survey (BOSS) and describe the correction procedure we have developed and applied to this sample. Calibration errors are primarily due to atmospheric differential refraction and guiding offsets during each exposure. The corrections potentially reduce the systematics for any studies of BOSS quasars, including the measurement of baryon acoustic oscillations using the Lyman-$\alpha$ forest. Our model suggests that, on average, the observed quasar flux in BOSS is overestimated by $\sim 19\%$ at 3600~\AA\ and underestimated by $\sim 24\%$ at 10,000~\AA. Our corrections for the entire BOSS quasar sample are publicly available. 
\end{abstract}


\section{Introduction}

One of the main goals of the third generation of the Sloan Digital Sky Survey \citep[SDSS-III;][]{2011AJ....142...72E} Baryon Oscillation Spectroscopic Survey \citep[BOSS;][]{2013AJ....145...10D} is to measure the baryon acoustic oscillation (BAO) scale in the Lyman-$\alpha$ forest from observations of high-redshift ($z > 2$) quasars \citep{2013A&A...552A..96B, 2013JCAP...04..026S, 2013JCAP...03..024K, 2014JCAP...05..027F, 2015A&A...574A..59D}. 
The transmitted flux fraction in the forest provides a measure of the neutral hydrogen density \citep{1965ApJ...142.1633G, 1998ARA&A..36..267R, 2009RvMP...81.1405M} along the line of sight that can be used to infer the clustering of the underlying dark matter distribution \citep{1994ApJ...437L...9C, 1995ApJ...452...90B, 1995ApJ...453L..57Z, 1996ApJ...457L..51H, 1996ApJ...471..582M, 1997ApJ...479..523B, 1997ApJ...486..599H, 1998MNRAS.301..478T}. 
Measurements of the BAO scale constrain the expansion history of the universe and can be used to infer the characteristics of dark energy \citep{2013PhR...530...87W}. 

During the period 2009-14, BOSS observed 294,512 quasars from a total sky area of 10,400 square degrees. 
As a result, the survey contains the largest sample of spectroscopic quasar observations to date and enables an unprecedented view into the multiple areas of quasar science, e.g. clustering of quasars \citep{2012MNRAS.424..933W}, quasar luminosity function \citep{2013A&A...551A..29P, 2013ApJ...773...14R, 2013ApJ...768..105M}, and variability properties of broad absorption lines in quasar spectra caused by high-velocity outflows \citep{2012ApJ...757..114F, 2013ApJ...777..168F, 2013MNRAS.434..222H, 2013ApJ...768...38V, 2014ApJ...791...88F}.

One potential source of systematics for any study of BOSS quasars is spectrophotometric calibration errors. 
Calibration errors in BOSS are larger than for SDSS-I \citep{2000AJ....120.1579Y} due to a design tradeoff that improves throughput in the Lyman-$\alpha$ forest of quasar spectra. 
In this work, we describe the dominant source of these errors in BOSS spectra and our procedure for reducing them. 
The miscalibration of BOSS spectra, on average, accounts for a $\sim 19\%$ excess at 3600~\AA\ and a $\sim 24\%$ decrement at 10,000~\AA\ with a smooth transition between (see \fig{correction-summary}). 

This paper is organized as follows. 
In \refsec{data}, we describe the SDSS data samples used in this analysis, along with the relevant charactistics of the survey design. 
We describe our methods for modeling and correcting the spectrophotometric calibration errors present in BOSS quasar spectra in \refsec{methods}. 
In \refsec{validation}, we evaluate our corrections using observations from a BOSS ancillary program that contains offset spectrophotometric standard stars, repeat observations of quasars observed in both SDSS-I and BOSS, and a sample of stellar contaminants in the BOSS quasar sample. 
Finally, we conclude with a discussion of our findings in \refsec{discussion}.

\section{Data Samples} \label{sec:data}

We use data from the \texttt{v5\_7\_0} BOSS spectroscopic pipeline \citep{2012AJ....144..144B} processing of the SDSS Data Release 12 \citep{2015arXiv150100963A}\footnote{\url{http://www.sdss.org/dr12/}}. 
There are two primary classes of targets used in this work: spectrophotometric standard stars (main sequence F stars used for calibration) and quasars; see \citet{2013AJ....145...10D} and \citet{2012ApJS..199....3R}, respectively, for descriptions of the target selection for these samples.  

The BOSS double spectrograph mounted at the Cassegrain focus of the 2.5-m SDSS telescope \citep{2006AJ....131.2332G} located at Apache Point Observatory\footnote{\url{http://www.apo.nmsu.edu/}} (APO) simultaneously records 1000 spectra over a $3^\circ$ field of view. The light from each target is captured by an optical fiber plugged into an aluminum plate at the focal plane and transported to either one of two spectrographs for analysis \citep{2013AJ....146...32S}. The circular focal plane hole for each 2 arcsecond ($120 \mu m$) diameter fiber is predrilled according to each target's sky position at a nominal observing time.
Each BOSS plate includes 20 spectrophotometric standard star targets which are used for spectrophotometric calibration. The calibrations are derived by fitting stellar spectrum models to the recorded spectra for these targets. 
Spectrophotometric calibration errors in BOSS quasar spectra are primarily due to offsets in fiber hole positioning between quasar targets and spectrophotometric standard stars that are intentionally introduced to improve the signal to noise ratio of the Lyman-$\alpha$ forest region of high-redshift quasars.
Additionally, there are 16 ``guide'' stars for each plate which are each observed with coherent fiber bundles and used for guiding the telescope during exposures; see \citet{2013AJ....145...10D} for more details. 

The BOSS DR12 sample contains a total of 487,276 targets with focal plane offsets, distributed between 2,377 observations of 2,340 plates. 
Of those targets, the BOSS data processing pipeline \citep{2012AJ....144..144B} identified 284,085 as quasars and 159,886 as stars. 
These stars are often referred to as ``failed quasars'', since they were targeted as quasars due to their photometric similarities \citep{2012ApJS..199....3R}. 
\tab{ntargets} summarizes the different target samples.

\begin{table*}
\centering
\begin{tabular}{l ccc rr }
    \hline
    Target Sample  & \textsc{lambda\_eff (\AA)} & \textsc{objtype} & \textsc{class} & $N_\mathrm{DR12}$ & $N_\mathrm{validation}$ \\
    \hline
    Offset targets   & \textsc{4000} & -                          & -             & 487,276 & 4,104  \\
    Quasars          & \textsc{4000} & \textsc{qso}               & \textsc{qso}  & 284,085 & 1,737  \\
    Failed quasars   & \textsc{4000} & \textsc{qso}               & \textsc{star} & 159,886 & 1,049  \\
    Spec. standards  & \textsc{5400} & \textsc{spectrophoto\_std} & \textsc{star} &  49,635 &   400  \\
    Offset standards & \textsc{4000} & -                          & \textsc{star} &   1,770 &   486  \\
    \hline
\end{tabular}
\caption{Target samples used in this work. For the DR12 samples, we include any potentially useful spectra, only requring that the fiber was plugged (i.e., that the \textsc{unplugged} bit of the \textsc{zwarning} bitmask is not set). For the validation samples, we additionally require that the spectra have no known problems (\textsc{zwarning} = 0) and that the spectra are from plates in the validation sample (\textsc{plate} keyword matches a plate in the validation set). The offset standards are identified via the \textsc{qso\_std} bit of the \textsc{ancillary\_target2} bitmask. 
\label{tab:ntargets}}
\end{table*}

In addition to the primary BOSS spectrophotometric standard stars, a small subset of observations contain a second sample of spectrophotometric standard stars which were selected and designed to have the same offsets in the focal plane as the BOSS quasar targets. 
This sample was collected as part of an ancillary program \citep{2013AJ....145...10D, 2015arXiv150100963A}. 
These stars were selected and visually inspected to ensure a uniform distribution across the focal plane, in a manner similar to normal spectrophotometric standard stars targeted in BOSS. 
The algorithm for photometric selection applied to this sample was identical to the algorithm for primary spectrophotometric standard stars in BOSS.

Spectrophotometric standard stars offset in the focal plane are identified by bit 20 of the \textsc{ancillary\_target2} keyword in DR12 data model\footnote{\url{http://www.sdss.org/dr12/algorithms/bitmasks/}}. 
We will refer to these objects as ``offset standards''. 
There are 1,770 offset standard star targets confirmed as stars distributed between 161 observations. 
Of those observations, 79 contain at least 10 offset standards in total, and 20 contain at least 10 offset standards per spectrograph. 

As part of this ancillary program, we modified the BOSS data processing pipeline to use the offset standards for spectrophotometric calibration, instead of the normal spectrophotometric standard stars. 
This sample provides a vital cross-check of the throughput correction model described below. 
In particular, we use the sample of 20 plates with at least 10 offset standards in each spectrograph for validation tests. 
In \tab{ntargets}, the column labeled $N_\mathrm{validation}$ indicates the number of targets in the validation sample for each of the relevant target samples. 
We refer to the sample of self-calibrated offset standard stars as ``offset standards''. 

The design wavelength $\lambda_i$ for a target is encoded in the \textsc{lambda\_eff} quantity of a \texttt{plateDesign} file and is propagated to a target's corresponding entry in \texttt{spPlate} and \texttt{spAll} files in the DR12 data model\footnote{\url{http://data.sdss3.org/datamodel/}}. 
The design hour angle $h_0$ for each plate is encoded in the \textsc{ha} quantity of its \texttt{plateHoles} file, but is not propagated to \texttt{spPlate} or \texttt{spAll}. 
We calculate the design altitude $a_0$ using the design hour angle and the central sky position of a plate.

\begin{figure}[ht]
\centering
\includegraphics[width=\columnwidth]{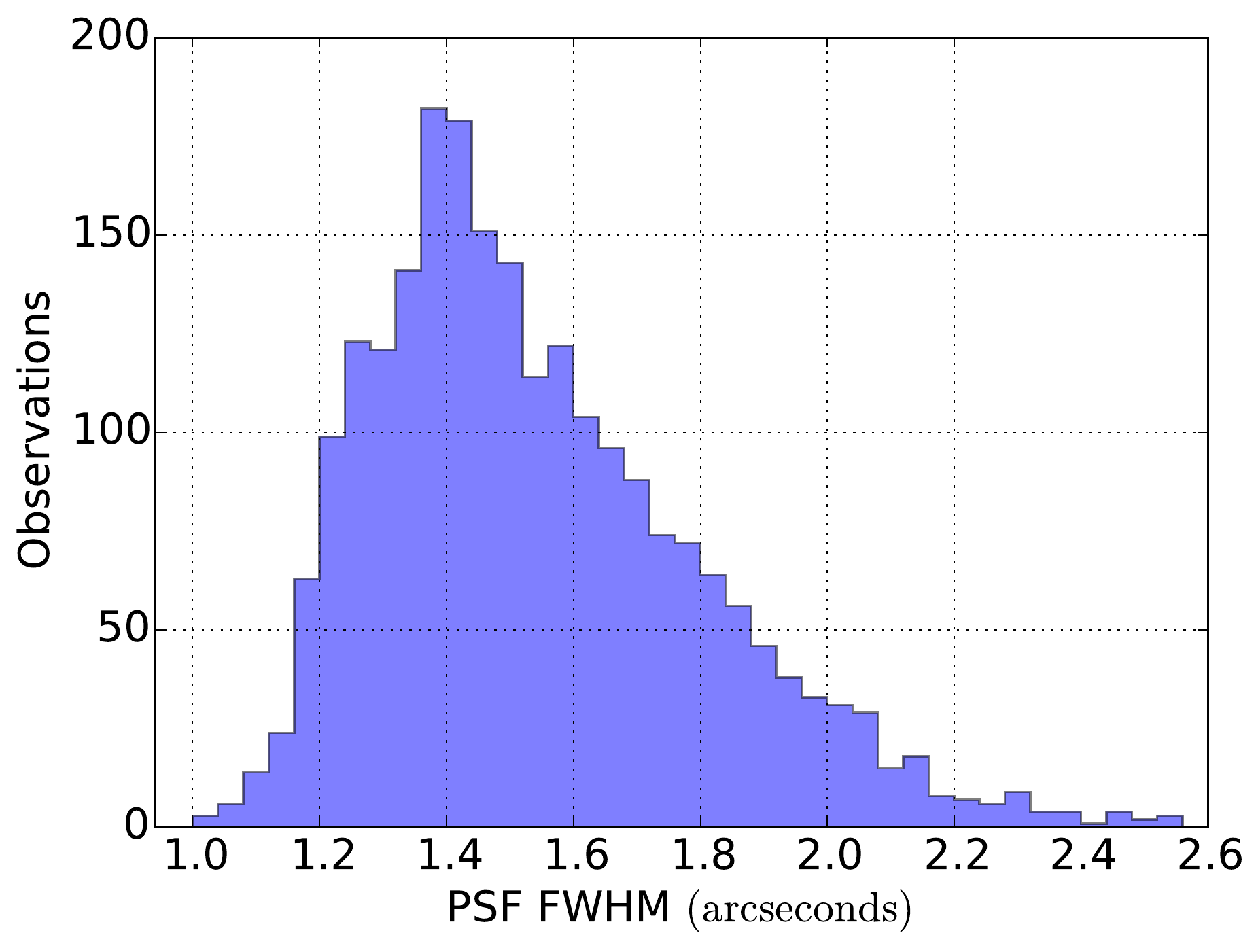}
\caption{Distribution of focal plane PSF FWHM, as indicated by the \textsc{seeing50} keyword, for DR12 observations. A total of 79 observations are missing this information and omitted here. The mean (median) FWHM is 1.54 arcseconds (1.49 arcseconds).}
\label{fig:psf-dist}
\end{figure}

\begin{figure}[ht]
\centering
\includegraphics[width=\columnwidth]{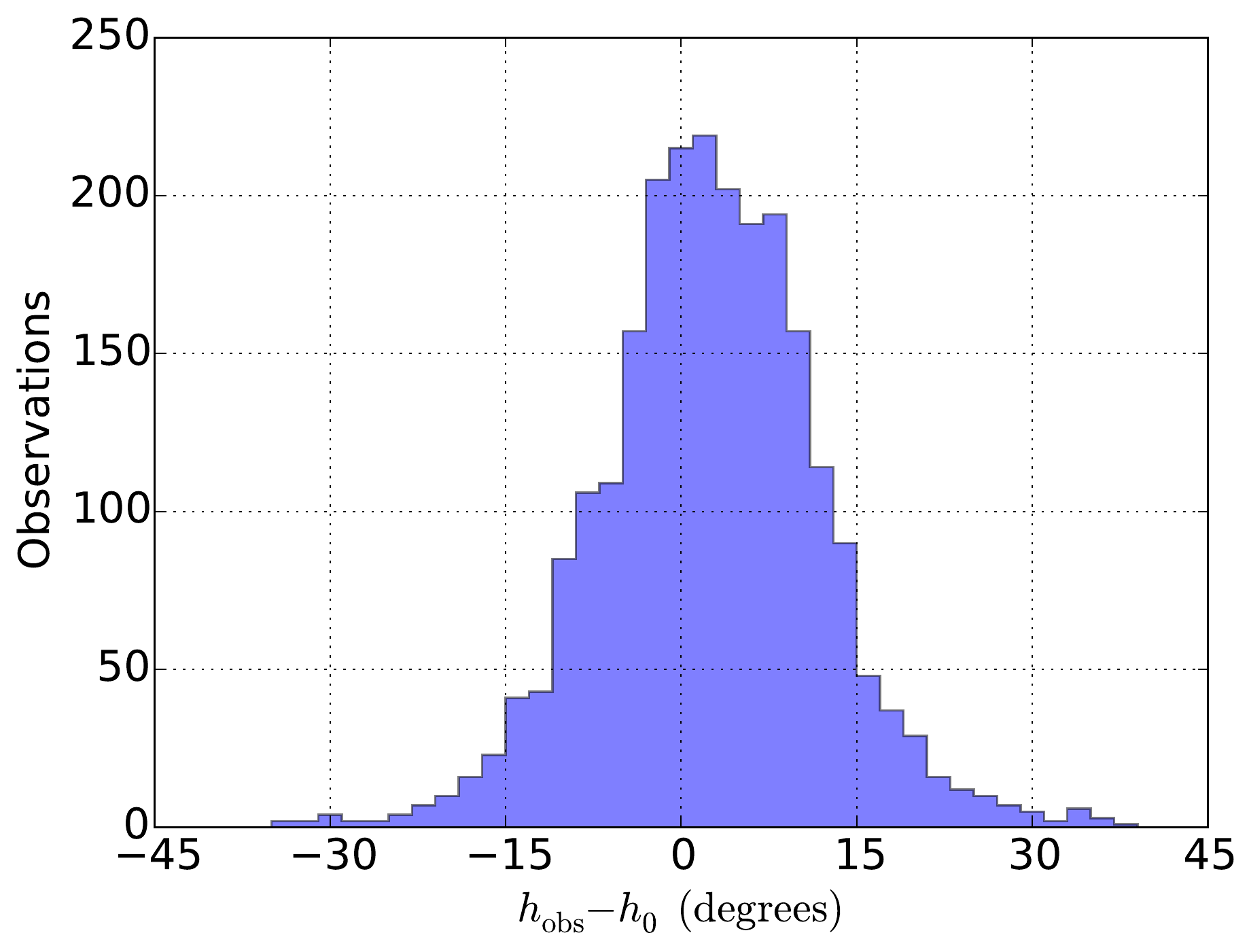}
\caption{Distribution of $h_\textrm{obs} - h_0$ for DR12 observations. The mean and RMS are $2.5^\circ$ and $9.4^\circ$, respectively.}
\label{fig:dha-dist}
\end{figure}

The distribution of point-spread function (PSF) full-width half-maxima (FWHM) for all observations in DR12 is shown in \fig{psf-dist}. 
For each observation, we summarize the PSF FWHM using an unweighted mean calculated from the individual exposures of the observation. 
The PSF FWHM is available via the \textsc{seeing50} keyword in \texttt{spCFrame} files for each exposure and is estimated from guide star images. 
Similarly, we calculate the mean hour angle for an observation $h_\textrm{obs}$ and the mean altitude $a_\textrm{obs}$ of the individual exposures. 
We calculate the observing hour angle from the midpoint of the times stored in \textsc{taibeg} and \textsc{taiend} keywords in \texttt{spCFrame} files for each exposure. We display the distribution of $h_\mathrm{obs}$ for plates relative to their design observing hour angle $h_0$ in \fig{dha-dist}.

The BOSS data processing pipeline combines individual 15-minute exposures, typically about four, into a single co-added spectrum. 
Ideally, the individual exposures should be corrected prior to co-addition, however, few BOSS analyses currently make use of individual exposures. 
Instead, we calculate a correction for co-added data using the mean PSF FWHM and observing hour angle of the exposures for an observation. 
There were 79 plates observed during the beginning of the survey which do not have PSF FWHM data recorded; for those observations, we use the sample median as an estimate for the PSF FWHM. 

\section{Methods} \label{sec:methods}

\subsection{Fiber illumination}

\begin{figure}[ht]
\centering
\includegraphics[width=\columnwidth]{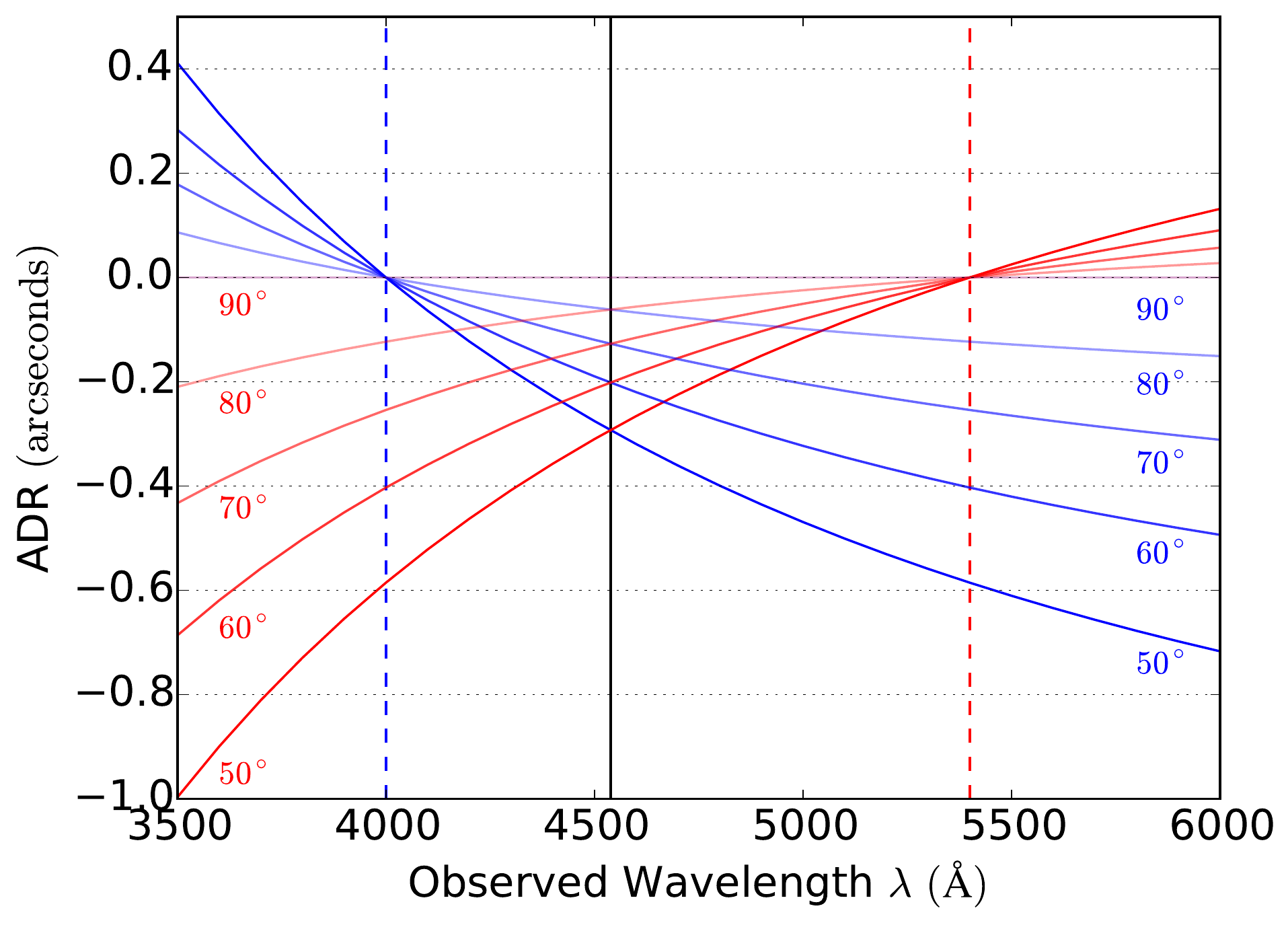}
\caption{Differential refraction of light in the atmosphere. The blue (red) curves show the differential refraction of light relative to 4000~\AA\ (5400~\AA). Lines show observing altitudes of $50^\circ$, $60^\circ$, $70^\circ$, $80^\circ$, and $90^\circ$. At $90^\circ$, the differential refraction is zero in both cases. The blue and red vertical dashed lines indicate the two relevant design waveslengths, 4000~\AA\ and 5400~\AA, respectively. The black vertical solid line indicates the interesection wavelength which is essentially independent of observing altitude.}
\label{fig:adr}
\end{figure}

The nominal transformation from sky coordinates $(\alpha,\delta)$ to focal-plane coordinates $\textbf{r} = (x,y)$ depends on:
\begin{itemize}
\item The sky coordinates $(\alpha_0,\delta_0)$ of the focal plane origin $\textbf{r} = 0$.
\item The wavelength $\lambda$ of incident light.
\item The time of the observation, expressed as the local hour angle $h$ for the right ascension $\alpha_0$.
\end{itemize}
The last two are due to wavelength-dependent refraction through the atmosphere with a time-dependent angle of incidence equal to the telescope altitude. We assume constant nominal conditions for atmospheric temperature and pressure for all refraction calculations. The atmospheric differential refraction (ADR) angle between 4000~\AA\ and 5400~\AA\ is about 0.5 arcseconds (the significance of these wavelengths is explained below), which is comparable to the 2 arcsecond diameter of BOSS fibers and the typical point-spread function (PSF) full-width half-maximum (FWHM) during DR12 observations of 1.5 arcseconds. In \fig{adr}, we show differential refraction angles relative to both 4000~\AA\ (blue) and 5400~\AA\ (red) light at various observing altitudes. The magnitude of ADR decreases at higher observing altitudes.

Each BOSS plate is designed for a specific pointing $(\alpha_0,\delta_0)$ at a nominal design hour angle $h_0$. The nominal orientation of the focal-plane coordinate system at hour angle $h$ is fixed such that the $+\hat{y}$ direction is aligned with increasing $\delta$ and $+\hat{x}$ is pointing eastwards. Photons at wavelength $\lambda$ observed at time $h$ from a target $i$ at sky coordinates $(\alpha_i,\delta_i)$ then have a PSF centered at the focal plane position
\begin{equation}
\textbf{r}_i(\lambda,h) \equiv \textbf{r}(\lambda,h; \alpha_i,\delta_i,\alpha_0,\delta_0) \; .
\end{equation}
The fiber hole for each target $i$ is positioned at $\textbf{r}(\lambda_i,h_0)$ so that the the target's nominal design wavelength $\lambda_i$ is centered on the fiber, and therefore has the maximum possible throughput, at time $h_0$.  Any departures from $\lambda = \lambda_i$ or $h = h_0$ will therefore introduce a fiber centering offset
\begin{equation}
d_i(\lambda,\lambda_i,h) \equiv \left| \textbf{r}_i(\lambda,h) - \textbf{r}(\lambda_i,h_0) \right|
\end{equation}
and a correspondingly reduced throughput that we calculate below. 

\begin{figure}[ht]
\centering
\includegraphics[width=\columnwidth]{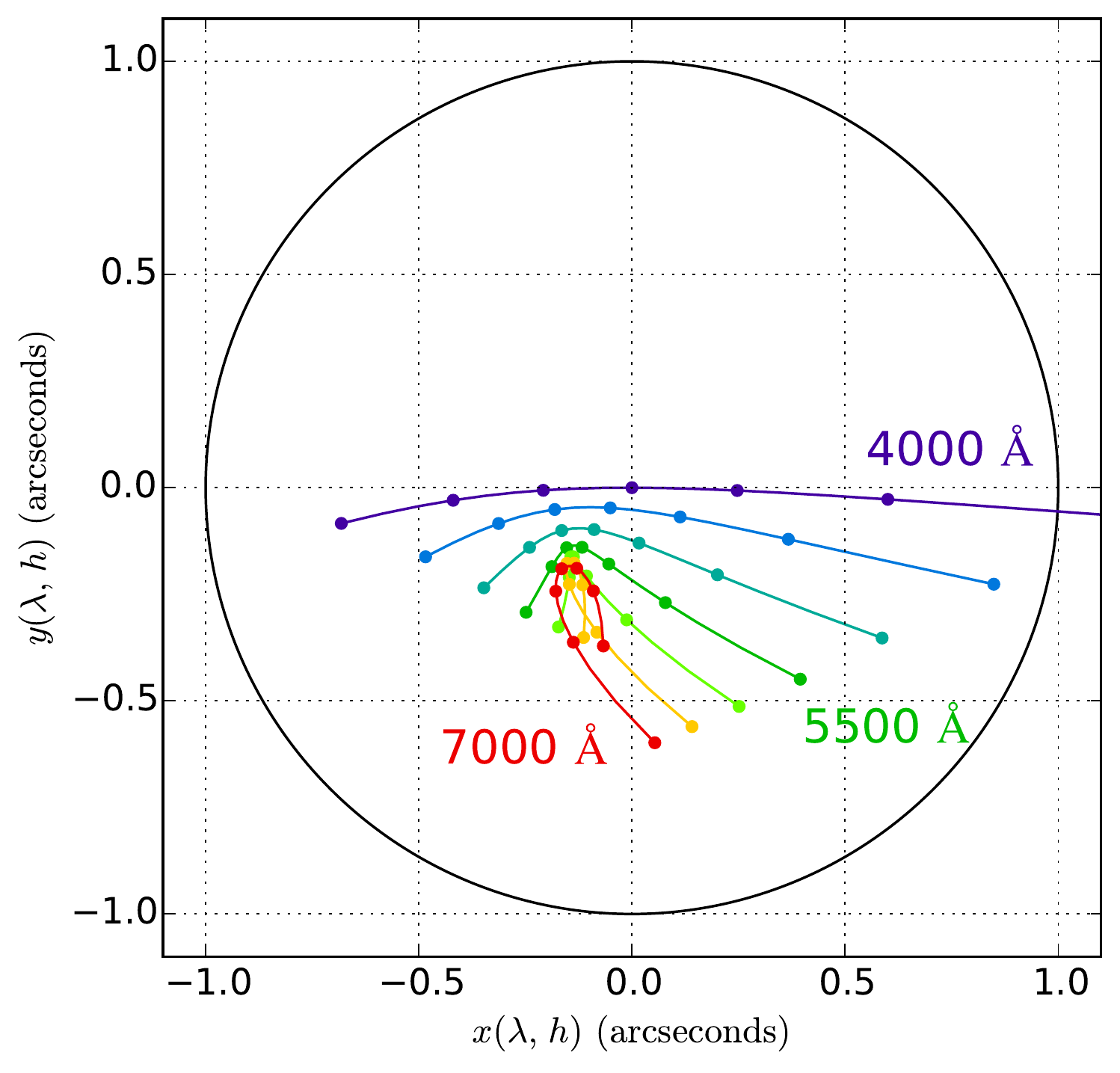}
\caption{Target centroid position relative to a fixed fiber hole as a function of wavelength $\lambda$ and observing hour angle $h$, as predicted for an ideal guiding model. The black circle represents a fiber hole (2 arcsecond diameter). The colored lines indicate the target's centroid position for specific wavelengths of incident light (4000~\AA\ to 7000~\AA\ in 500~\AA\ intervals) as a function of observing hour angle $h$. The observing hour angle window shown here is $\pm$ 3 hours from the design hour angle $h_0$ of the plate. The spacing between points is 1 hour. The fiber hole is positioned such that $\lambda_i = 4000$~\AA\ light is centered at the design hour angle $h_0$.}
\label{fig:fiber-hole}
\end{figure}

In practice, the majority of fibers are positioned using the same central wavelength $\lambda_i = 5400$~\AA. The exception are fibers assigned to high-redshift quasar targets, which use a bluer $\lambda_i = 4000$~\AA\ in order to improve the throughput (and hence also signal-to-noise ratio) in the Lyman-$\alpha$ forest region ($\lambda_{\mathrm{Ly}\alpha}(1+z)$ = 4000~\AA\ at $z = 2.29$). \fig{fiber-hole} shows an example of $\textbf{r}^\ast_i(\lambda,h)$ trajectories (the asterisked quantity is explained in the following section) plotted for a single quasar target with $\lambda_i = 4000$~\AA, covering a range of wavelengths $\lambda$ and hour angles $h$. The origin of the calibration errors that we seek to fix in this paper is this difference of design wavelengths. Focal-plane offsets were first introduced in the SDSS-III BOSS survey, so quasar spectra recorded during SDSS-I do not have these calibration errors.

\begin{figure}[ht]
\centering
\includegraphics[width=\columnwidth]{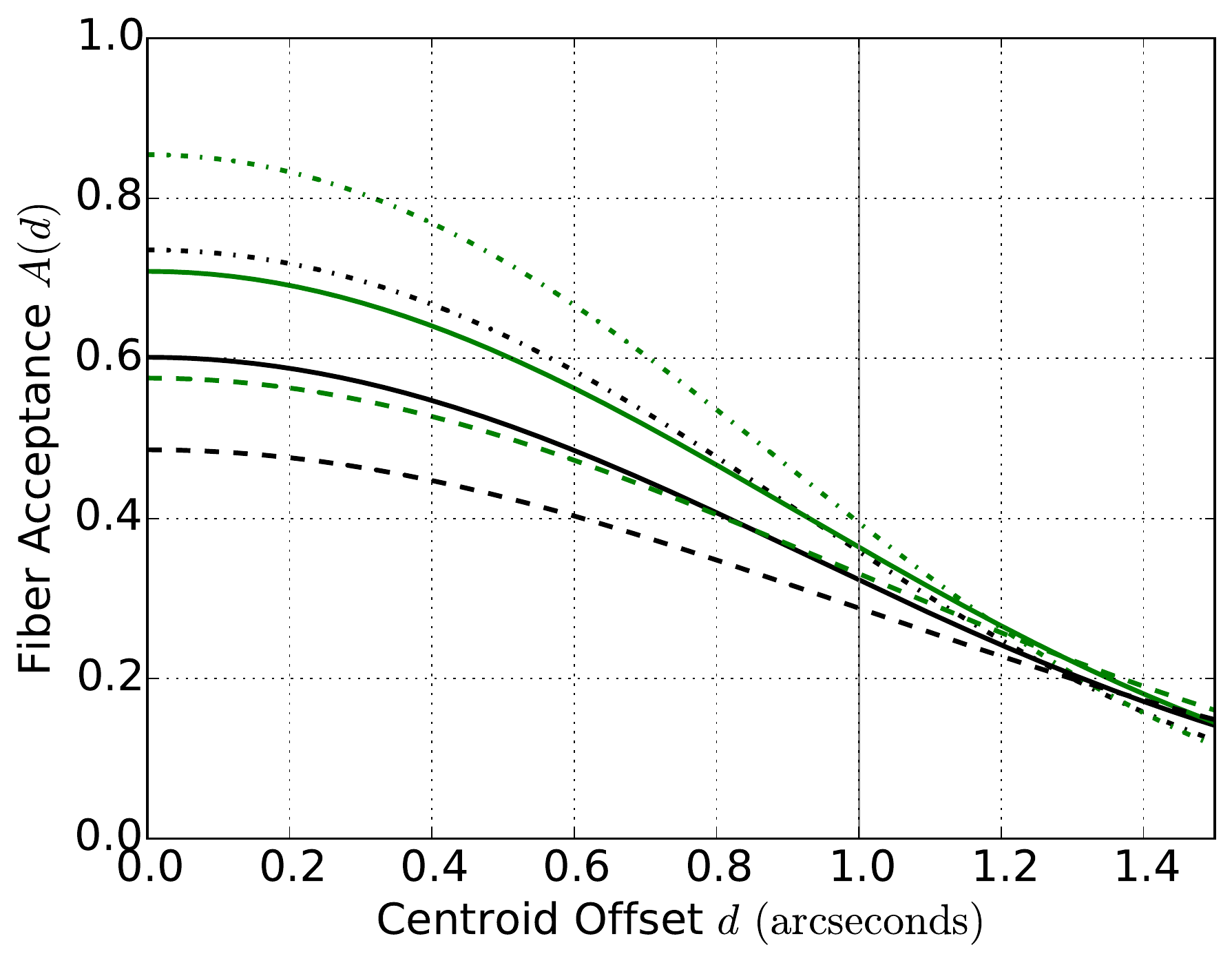}
\caption{Acceptance fraction as a function of offset distance for typical PSF sizes. The dotted, solid, and dashed curves respectively correspond to acceptance fractions calculated using PSF FWHM values of 1.2, 1.5, and 1.8 arcseconds. The green and black curves respectively correspond to Gaussian and Kolmogorov PSF shapes. The solid vertical line indicates the BOSS fiber aperture radius.}
\label{fig:acceptance_fractions}
\end{figure}

The overall system throughput for target $i$ includes a geometrical factor $A_i(\lambda,\lambda_i,h)$ that measures the fraction of incident light that enters its aperture. This fiber acceptance fraction depends on the PSF size and shape and its centroid offset $d_i(\lambda,h)$.  In the following, we assume that all PSFs have circular Gaussian profiles described by a standard deviation $\sigma$ that is independent of wavelength and constant during an exposure, so that
\begin{equation}
A_i(\lambda,\lambda_i,h) = A_{\text{Gauss}}(\sigma,d_i(\lambda,\lambda_i,h))
\end{equation}
with
\begin{equation}
A_{\text{Gauss}}(\sigma,d) \equiv \sigma^{-2} \int_0^{D/2}\, e^{-(d^2+r^2)/(2\sigma^2)} I_0(r d / \sigma^2)\, dr \; ,
\end{equation}
where $D = 2$ arcseconds is the BOSS fiber diameter and $I_0$ is a modified Bessel function of the first kind. We show later that our corrections are relatively insensitive to these assumptions since they involve fiber-acceptance ratios. \fig{psf-dist} shows the distribution of PSF sizes measured for DR12 observations and \fig{acceptance_fractions} shows examples of acceptance fractions calculated under different assumptions.

\subsection{Telescope guiding}

BOSS exposures are typically 15 minutes long, during which time an active telescope guiding loop uses 16 bright stars in the focal plane to maintain the pointing towards $(\alpha_0,\delta_0)$. Without any additional adjustments, the variation of $d_i(\lambda,\lambda_i,h)$ during the exposure would noticeably reduce target throughputs relative to their optimum values at $h_0$ where $d_i(\lambda_i,\lambda_i,h_0) = 0$. Therefore the guide loop also makes small adjustments to the focal plane position $\delta \textbf{r}$, radial scale $s$, and rotation $\theta$ during each exposure, which modify the transformation from sky to focal-plane coordinates according to
\begin{equation}
\textbf{r} \rightarrow G(s,\theta)\cdot \textbf{r} + \delta \textbf{r} \quad , \quad G(s,\theta) \equiv \left(1 + s\right) R(\theta)
\end{equation}
where $R$ is a 2D rotation matrix acting on focal-plane coordinates. The corresponding centering offset for fiber $i$ is then
\begin{multline}
d_i'(\lambda,\lambda_i,h) = \\
\left|\, G(s(h),\theta(h))\cdot \textbf{r}_i(\lambda,h) + \delta \textbf{r}(h) - \textbf{r}_i(\lambda_i,h_0)\,\right| \; ,
\end{multline}
where the prime denotes the effects of the guiding adjustments $\delta \textbf{r}(h)$, $s(h)$, and $\theta(h)$.

\begin{figure*}[ht]
\centering
\includegraphics[width=\textwidth]{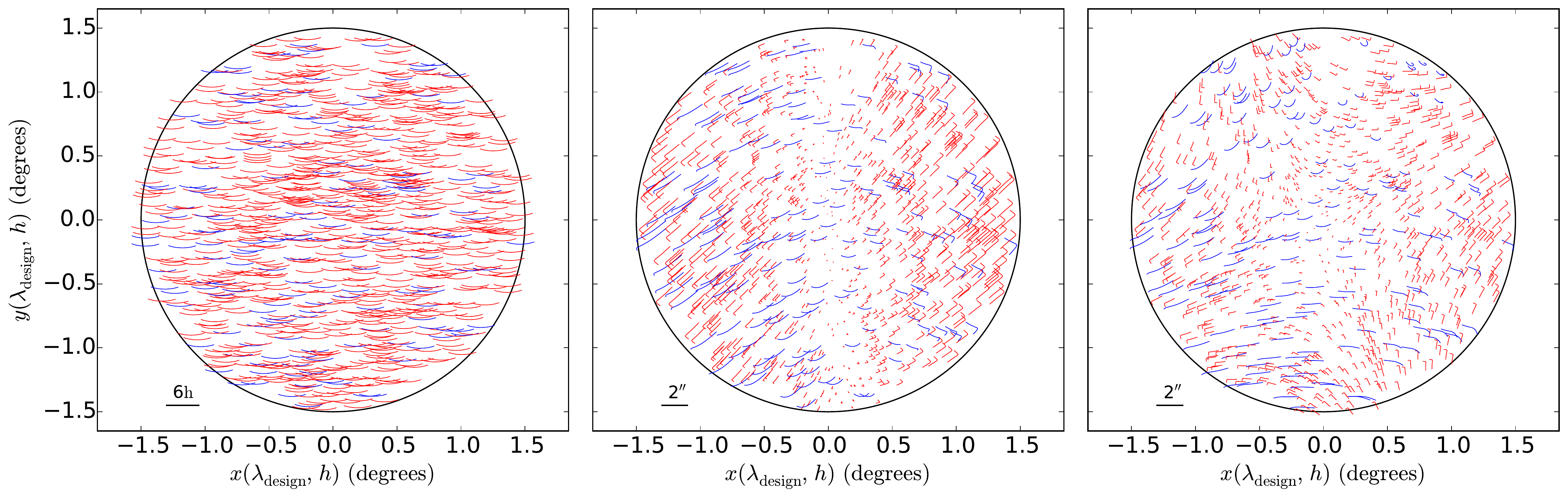}
\caption{The ideal guiding model for an example plate (\textsc{PLATE = 6114}). The black circle indicates the plate's $3^\circ$ field of view. The colored lines represent a target's $\lambda_i$ light path spanning a range of $\pm$ 3 hours from the design hour angle of the plate. Red (blue) lines indicate $\lambda_i = 5400$~\AA\ (4000~\AA) targets. The fiber centroid offset distances have been exagerated relative to the plate dimensions with the scale indicated in the bottom left of each panel.}
\label{fig:guiding-montage}
\end{figure*}

It is not feasible to reconstruct the actual history of guiding adjustments $\delta \textbf{r}(h)$, $s(h)$, and $\theta(h)$ during an exposure so we instead adopt an ideal guiding model, which assumes that the plate center tracks $(\alpha_0,\delta_0)$ optimally and that the guider makes continuous adjustments to minimize the chi-square figure of merit
\begin{equation}
\chi^2(s(h),\theta(h)) \equiv \sum_g d_g'(\lambda,\lambda_i,h)^2
\end{equation}
where the sum is over the 16 guide stars, indexed by $g$ and given equal weight, and the resulting adjustments optimize throughput at $\lambda_g = 5400$~\AA.  We denote the trajectories resulting from ideal guiding as $\textbf{r}_i^\ast(\lambda,h)$ and the corresponding centering offsets as 
$d_i^\ast(\lambda,\lambda_i,h)$. \fig{guiding-montage} shows an example of trajectories $\textbf{r}_i(\lambda_i,h)$ (left panel) and $\textbf{r}_i^\ast(\lambda_i,h)$ (right panel) at each target's design wavelength $\lambda_i$ for an entire plate. The middle panel shows trajectories when only the plate center $(\alpha_0,\delta_0)$ is tracked and no additional adjustments are made to track the guide stars. We find that, under the ideal guiding assumption, the translation, scale, and rotation degrees of freedom completely remove the monopole and dipole contributions to centering offsets for targets with $\lambda_i = \lambda_g = 5400$~\AA\ (shown in red), but lead to larger offsets with some residual monopole and dipole for the smaller set of quasar targets with $\lambda_i = 4000$~\AA\ (shown in blue).

\subsection{Spectrophotometric calibration}

The mean rate of detected photons from source $i$ at wavelength $\lambda$ and hour angle $h$ is related to the source's true spectral energy distribution (SED) $f_i(\lambda)$ by a fiber-dependent calibration factor $C$
\begin{equation}
\frac{dn_i}{dh}(\lambda,\lambda_i,h) = f_i(\lambda)\,C_i(\lambda,\lambda_i,h) \; ,
\end{equation}
which we split
\begin{equation}
C_i(\lambda,\lambda_i,h) = A_i^\ast(\lambda,\lambda_i,h) B_i(\lambda,h)
\end{equation}
into the fiber-acceptance fraction $A_i^\ast(\lambda,\lambda_i,h)$ assuming ideal guiding (denoted by the asterisk), and a term $B_i$ that includes all other contributions to the signal throughput but is independent of a target's design wavelength $\lambda_i$. The quantities that we actually measure from an exposure are the integrated photon counts
\begin{align}
n_i(\lambda,\lambda_i,h_{\text{obs}}) &= \int_{h_{\text{obs}}-\Delta h/2}^{h_{\text{obs}} + \Delta h/2}
\frac{dn_i}{dh}(\lambda,\lambda_i,h)\, dh \\
&\simeq \frac{dn_i}{dh}(\lambda,\lambda_i,h_{\text{obs}})\,\Delta h
\; ,
\end{align}
where $h_{\text{obs}}$ is the exposure midpoint hour angle and $\Delta h = 3.75^\circ$ (15 minutes) is the exposure duration. The variation of the integrand is sufficiently small over $\Delta h$ that we use the trapezoidal approximation to the integral in the following. \fig{dha-dist} shows the distribution of $h_{\text{obs}} - h_0$ for all DR12 observations.

The 20 spectrophotometric calibration targets on each plate, indexed by $c$, are chosen so that their true SED $f_c(\lambda)$ can be directly estimated as $\tilde{f}_c(\lambda)$ using a stellar-model fit to their observed photon counts $n_c(\lambda,\lambda_c,h_{\text{obs}})$. We use these SED estimates to provide estimates of the calibration factors $\tilde{C}_c(\lambda,\lambda_c,h_{\text{obs}})$ at 20 locations across each plate, and then interpolate these factors to the locations of every fiber, $\tilde{C}_i(\lambda,\lambda_i,h_{\text{obs}})$. We then estimate the SED of all targets on a plate as
\begin{equation}
\tilde{f}_i(\lambda) = \tilde{C}_i(\lambda,\lambda_i,h_{\text{obs}})^{-1}\, n_i(\lambda,\lambda_i,h_{\text{obs}}) / \Delta h \; .
\end{equation}

\begin{figure*}
\centering
\includegraphics[width=\textwidth]{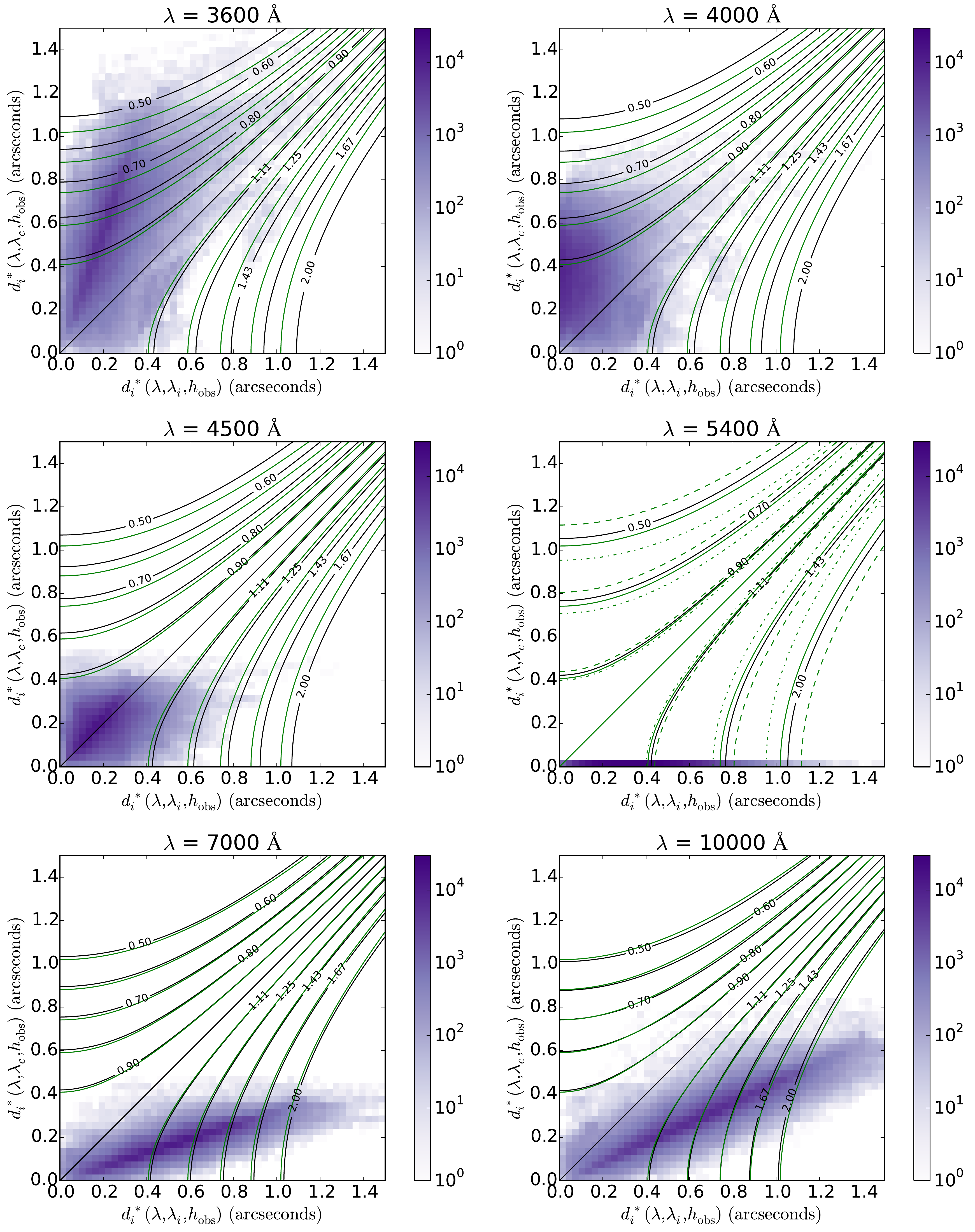}
\caption{2D histrogram of $d_i^\ast(\lambda,\lambda_i,h_\mathrm{obs})$ and $d_i^\ast(\lambda,\lambda_c,h_\mathrm{obs})$ focal plane position offsets relative to the fiber hole center for all offset targets. The color of each 2D bin in a panel corresponds to the number of entries as indicated by the panel's adjacent vertical color bar. The green and black contours represent levels of $R_i^\ast(\lambda)$ calculated assuming Gaussian and Kolmogorov PSF shapes, respectively. In the panel labeled $\lambda = 5400$~\AA\, the additional dotted, solid, and dashed contours represent levels of $R_i^\ast(\lambda)$ for PSF FWHM of 1.2, 1.5, and 1.8 arcseconds, respectively.}
\label{fig:offset-compare-montage}
\end{figure*}

A key point of this paper is that {\em the method used by the BOSS data processing pipeline to interpolate the standard-star calibrations $\tilde{C}_c$ to other targets in the focal plane does not account for differences in each target's design wavelength $\lambda_i$.} Since the calibration stars have $\lambda_c = 5400$~\AA, this only affects the quasar targets, which are designed with $\lambda_i = 4000$~\AA. Under the ideal guiding assumption, the SED mis-calibration of targets with $\lambda_i \ne \lambda_c$ is the ratio of acceptance fractions at the actual ($\lambda_i$) and assumed ($\lambda_c$) design wavelengths
\begin{equation}
R_i^\ast(\lambda) \equiv \frac{A_i^\ast(\lambda,\lambda_c,h_{\text{obs}})}{A_i^\ast(\lambda,\lambda_i,h_{\text{obs}})} \; ,
\end{equation}
and our improved calibration consists of scaling the pipeline SED estimates by this factor
\begin{equation}
\tilde{f}_i(\lambda) \rightarrow R_i^\ast(\lambda)\,\tilde{f}_i(\lambda) \; .
\end{equation}
To summarize, the corrections $R_i^\ast(\lambda)$ that we implement depend on:
\begin{itemize}
\item The actual fiber centering offsets $d_i^\ast(\lambda,\lambda_i,h_{\text{obs}})$ and the assumed offsets $d_i^\ast(\lambda,\lambda_c,h_{\text{obs}})$ at the exposure midpoint $h_{\text{obs}}$ (both under ideal guiding conditions).
\item The design hour angle $h_0$, which determines the locations of each fiber on the focal plane.
\item The PSF size $\sigma$.
\end{itemize}
\fig{offset-compare-montage} shows scatter plots of $d_i^\ast(\lambda,\lambda_i,h_{\text{obs}})$ versus $d_i^\ast(\lambda,\lambda_c,h_{\text{obs}})$ for all offset targets in DR12 observations ($h_{\text{obs}}$ and $h_0$), for six different wavelengths $\lambda$, with contours of $R_i^\ast(\lambda)$ superimposed for three different PSF sizes $\sigma$. Additionally, we show contours of $R_i^\ast(\lambda)$ for a nominal PSF size assuming a Kolmogorov atmospheric turbulence model \citep{1961wptm.book.....T} which includes a wavelength dependent scaling of the nominal PSF FWHM $\propto \lambda^{-1/5}$ \citep{FRIED:66}.

\begin{figure*}[ht]
\centering
\includegraphics[width=\textwidth]{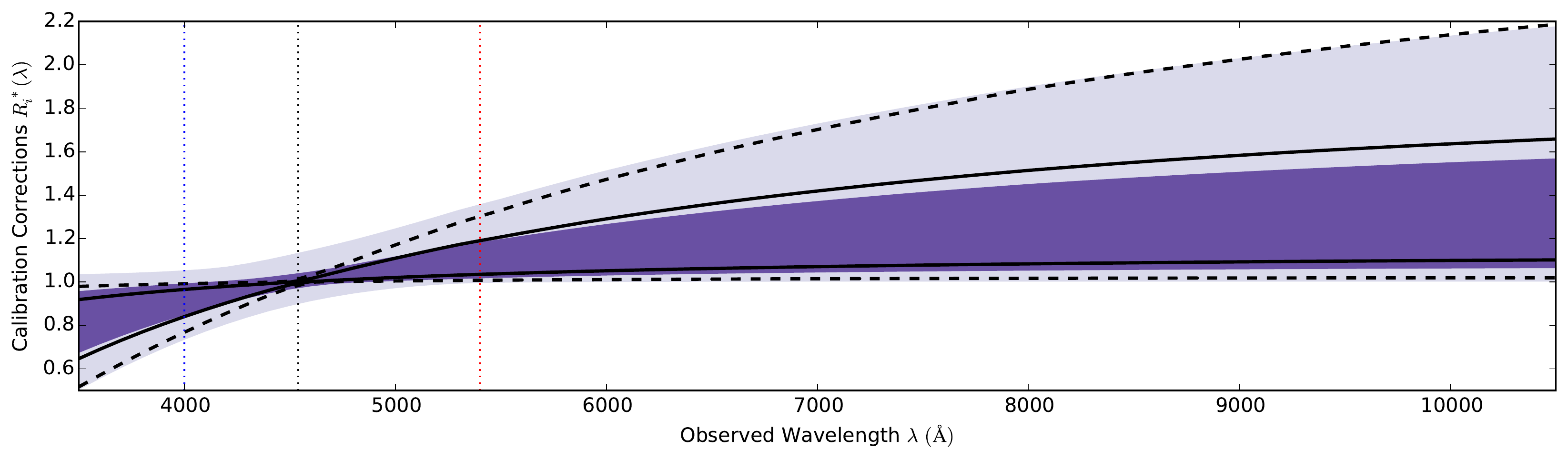}
\caption{Throughput correction summary. The shaded regions correspond to the the central 68\% and 95\% quantiles of the predicted correction at each wavelength. The black solid and dashed lines similarily correspond to the central 68\% and 95\% quantiles of $R_i^\ast(\lambda)$ for observations where $|h_\mathrm{obs} - h_0| < 1.25^\circ$ (5 minutes), approximately 10\% of all observations.}
\label{fig:correction-summary}
\end{figure*}

In \fig{correction-summary}, we show the central 68\% and 95\% quantiles of $R_i^\ast(\lambda)$ for all 487,276 offset targets in DR12. In general, the corrections $R_i^\ast(\lambda)$ are smooth, monotonic functions that cross unity near 4500~\AA (as expected, see \fig{adr}). Comparing with the subset of $R_i^\ast(\lambda)$ where $|h_\mathrm{obs} - h_0| < 1.25^\circ$ (5 minutes), approximately 10\% of all observations, we see that the vertical spread at the crossover $\lambda$ is caused by plates observed away from their design hour angle. On average, our model suggests that the observed quasar flux is overestimated by $\sim 19\%$ at and underestimated by $\sim 24\%$ near 3600~\AA\ and 10,000~\AA, respectively. 

\subsection{Correction Parametrization}

\begin{figure}
\centering
\includegraphics[width=\columnwidth]{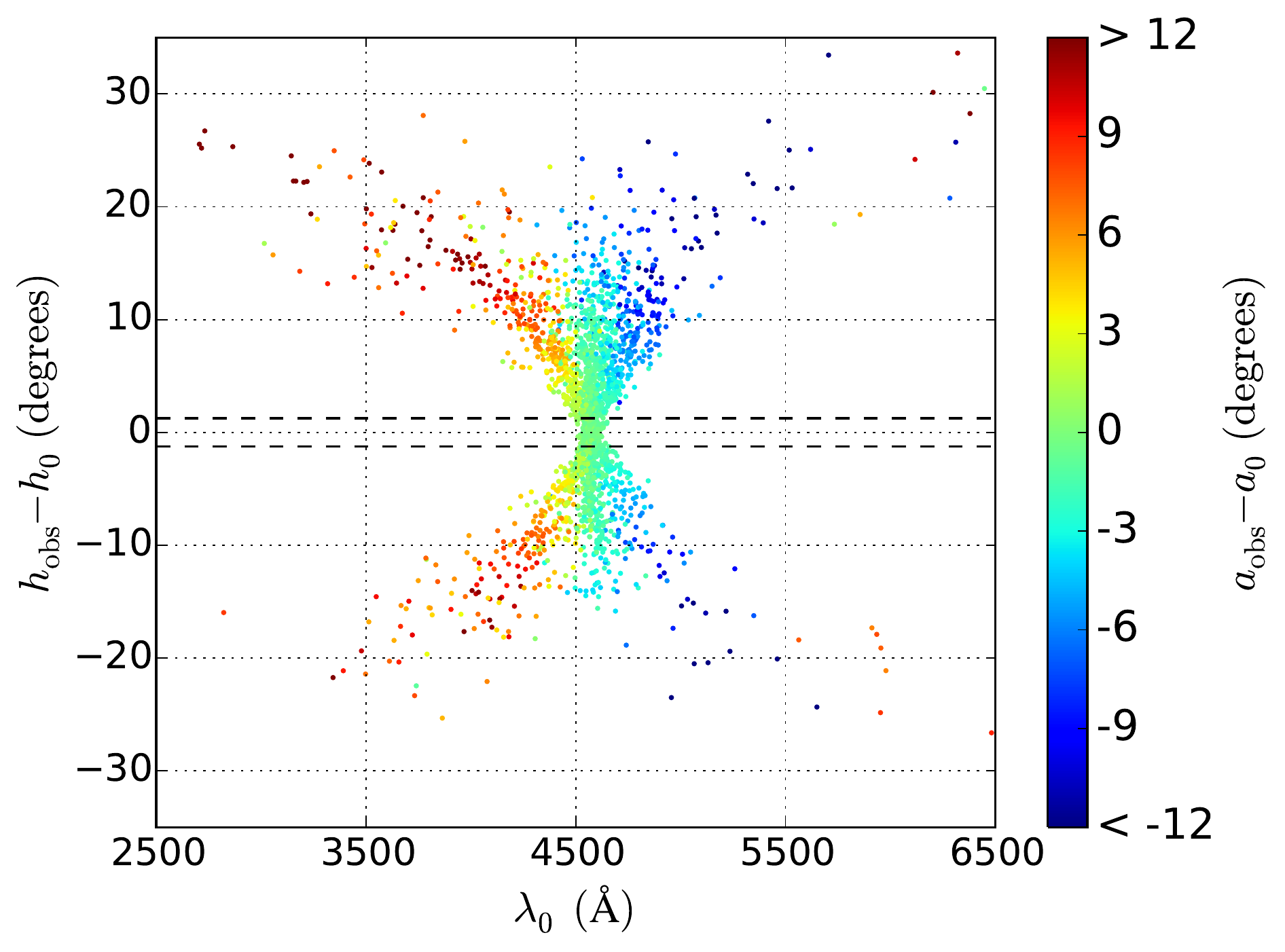}
\caption{Relationship between median fit parameter $\lambda_0$ and hour angle difference $(h_\textrm{obs} - h_0)$ for all 2377 DR12 plates with offset targets. Points are colored by the corresponding altitude difference $(a_\textrm{obs} - a_0)$ for the observation.}
\label{fig:fit-lambda0-scatter}
\end{figure}

\begin{figure}
\centering
\includegraphics[width=\columnwidth]{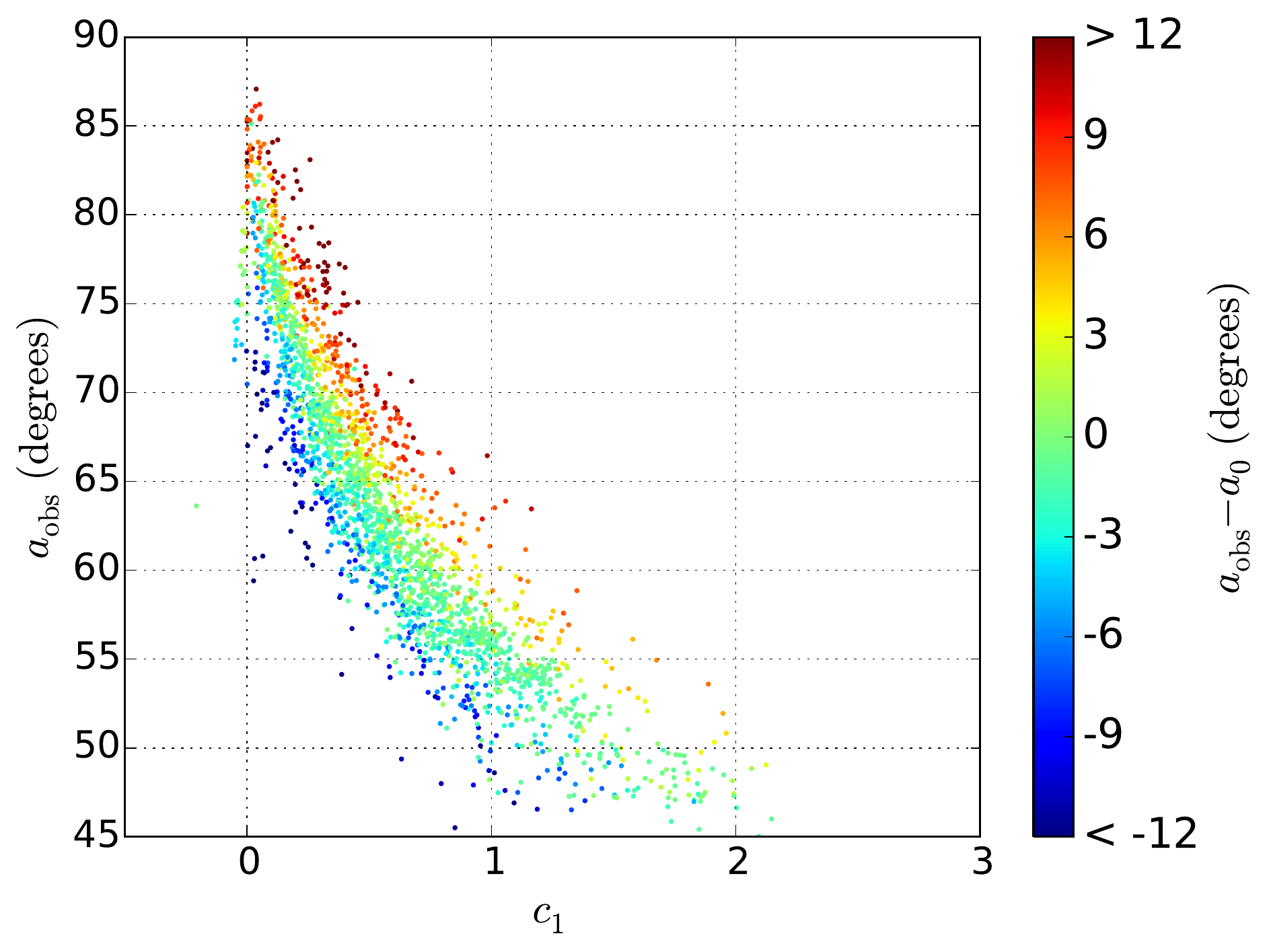}
\caption{Relationship between median fit parameter $c_1$ and observing altitude $a_\textrm{obs}$ for all 2377 DR12 plates with offset targets. Points are colored by the corresponding altitude difference $(a_\textrm{obs} - a_0)$ for the observation.}
\label{fig:fit-c1-scatter}
\end{figure}

\begin{figure}
\centering
\includegraphics[width=\columnwidth]{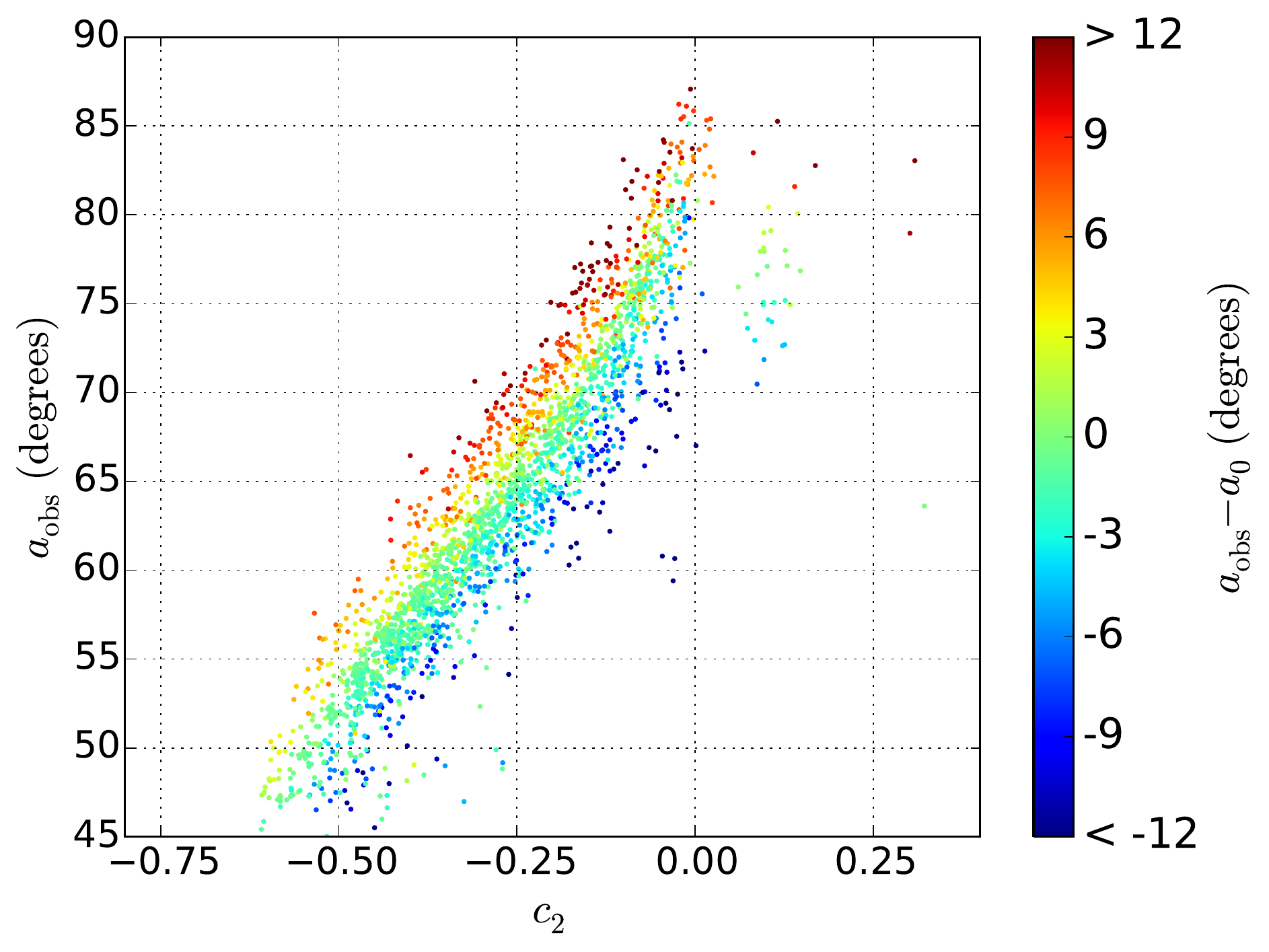}
\caption{Relationship between median fit parameter $c_2$ and observing altitude $a_\textrm{obs}$ for all 2377 DR12 plates with offset targets. Points are colored by the corresponding altitude difference $(a_\textrm{obs} - a_0)$ for the observation.}
\label{fig:fit-c2-scatter}
\end{figure}

\begin{figure}
\centering
\includegraphics[width=\columnwidth]{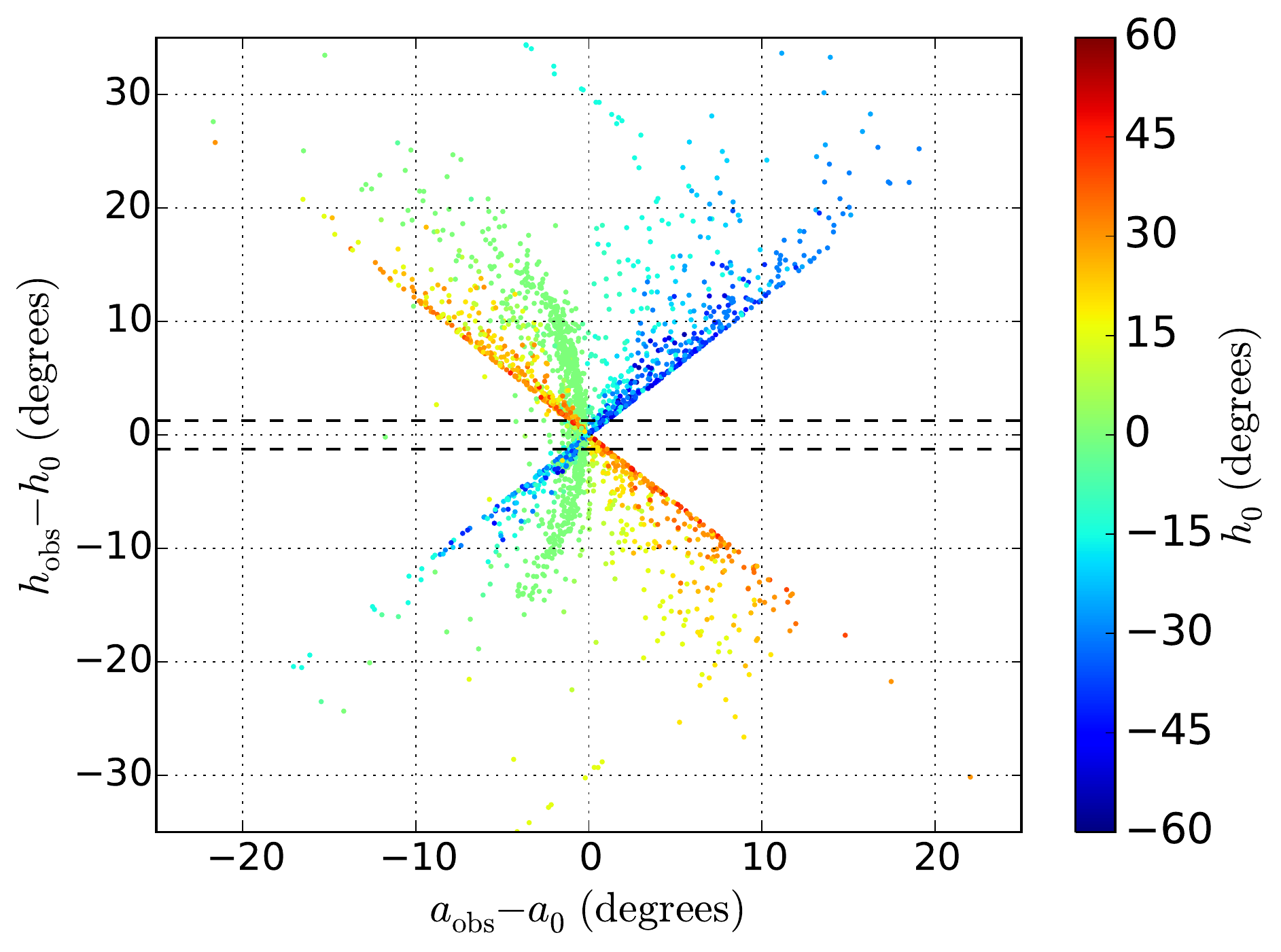}
\caption{Relationship between the hour angle difference $(h_\textrm{obs} - h_0)$ and altitude difference $(a_\textrm{obs} - a_0)$ for all 2377 DR12 plates with offset targets. Points are colored by the corresponding design hour angle $h_0$ for the observation.}
\label{fig:dalt-dha-scatter}
\end{figure}

In order to summarize the general properties of our correction model for all DR12 observations, we fit a parametrized model 
\begin{equation}
1 + c_1 \log \lambda/\lambda_0 + c_2 \left(\log \lambda/\lambda_0\right)^2
\end{equation}
to each throughput correction prediction $R_i^\ast(\lambda)$. Figures~\ref{fig:fit-lambda0-scatter}-\ref{fig:fit-c2-scatter} show the median best fit parameter per plate and its relationship to either $a_\textrm{obs}$ or $(h_\textrm{obs} - h_0)$. 
The points in each figure are colored by $(a_\textrm{obs} - a_0)$ for the corresponding observation. The $\lambda_0$ parameter, which corresponds to the crossover point of the multiplicative correction, is strongly correlated with $(a_\textrm{obs} - a_0)$, as seen in \fig{fit-lambda0-scatter}. 
The value of $c_1$ sets the overall size of the correction and is strongly correlated with the altitude of the observation, as seen in \fig{fit-c1-scatter}. \fig{fit-c2-scatter} shows a similar relationship betwen the magnitude $c_2$ and observing altitude. Observations at higher altitude tend to have smaller $c_1$ and $c_2$ values, which corresponds to a smaller correction.
For reference, we also display the relationship between $(h_\textrm{obs} - h_0)$ and $(a_\textrm{obs} - a_0)$ in \fig{dalt-dha-scatter}.

\section{Validation} \label{sec:validation}

\begin{figure*}
\centering
\includegraphics[width=\textwidth]{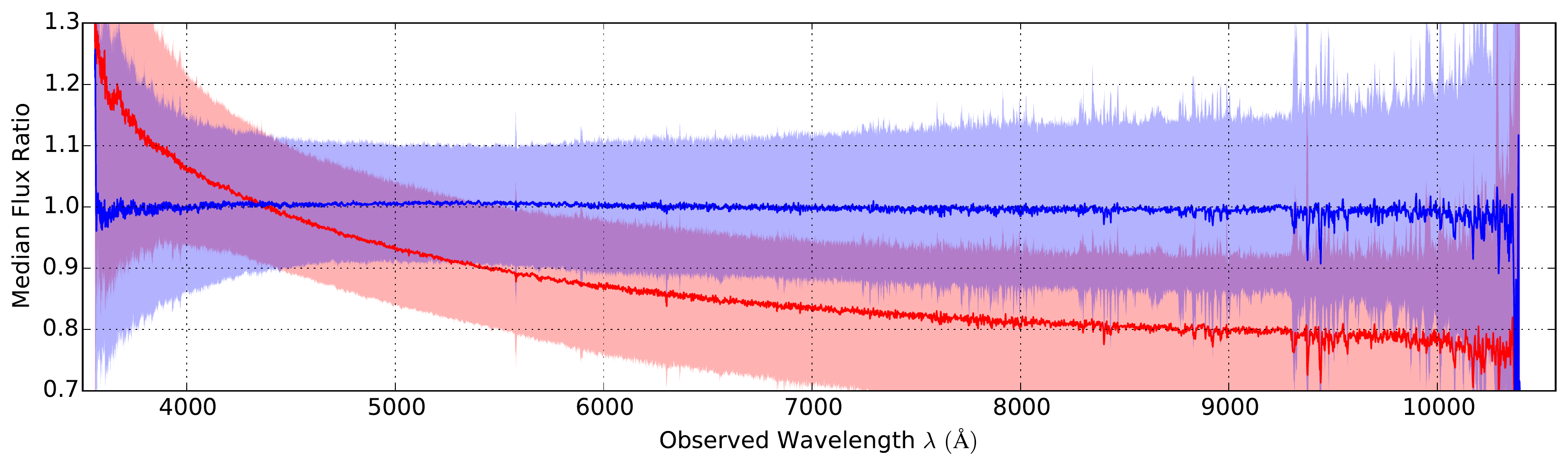}
\caption{Median flux ratios of failed quasar spectra from the 20 validation plates. The blue (red) curve shows the median flux ratio between corrected (uncorrected) BOSS spectra and their corresponding ``validation'' calibrated spectra. The shaded region corresponds to the 68\% level of the distribution in each wavelength bin. }
\label{fig:validation}
\end{figure*}

\begin{figure*}
\centering
\includegraphics[width=\textwidth]{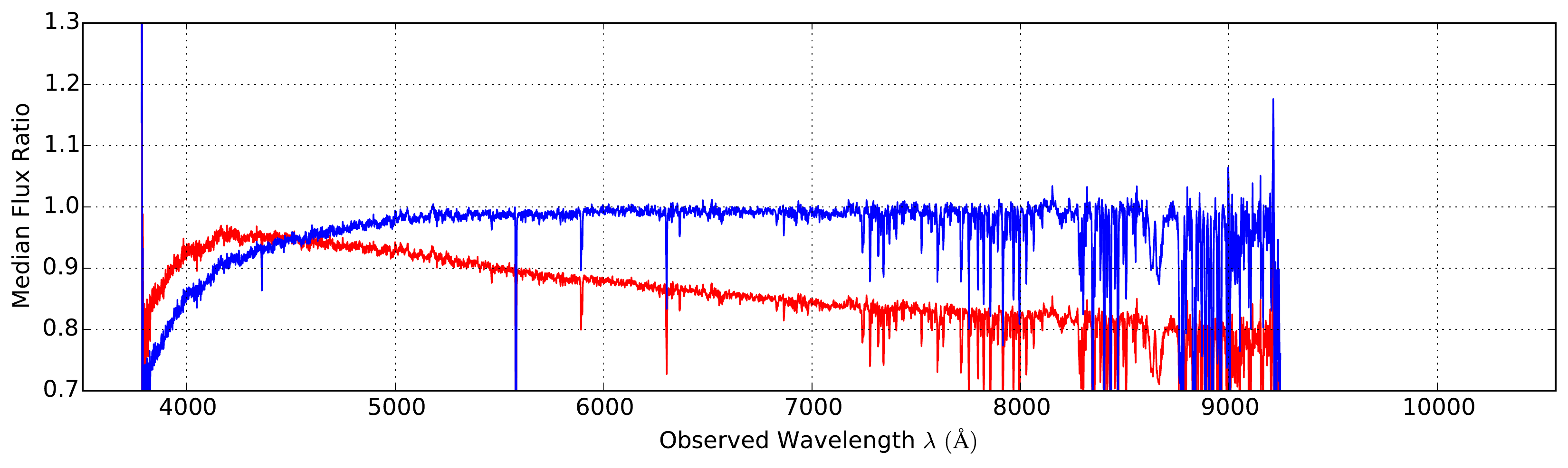}
\caption{Median flux ratios of quasar spectra that were observed in both SDSS-I and BOSS. The blue (red) curve shows the median flux ratio between corrected (uncorrected) BOSS spectra and their corresponding SDSS-I spectra. }
\label{fig:repeat-comparison}
\end{figure*}

In order to validate our model, we use observations that have been calibrated with offset standards, as described previously. In \fig{validation}, we display the median flux ratio of DR12 failed quasar spectra, with (blue) corrections and without (red) corrections, compared to their corresponding spectra from the ``validation'' reduction. The corrections remove a wavelength-dependent systematic bias in the distribution of BOSS quasar spectra. The vertical spread in each distribution indicates that there are other factors affecting the calibration but that they are independent of the effect described in this work. 
\fig{repeat-comparison} displays a similar comparison between a set of quasar targets with spectroscopic observations in both BOSS and SDSS-I. In this case, the SDSS-I observation is used for reference since focal plane offsets were not introduced prior to BOSS. We note a significant drop off below 5000~\AA but do not necessarily expect agreement between SDSS-I and BOSS because of changes to the spectrographs and the intrinsic variability of quasars on these timescales.

\begin{figure}
\centering
\includegraphics[width=\columnwidth]{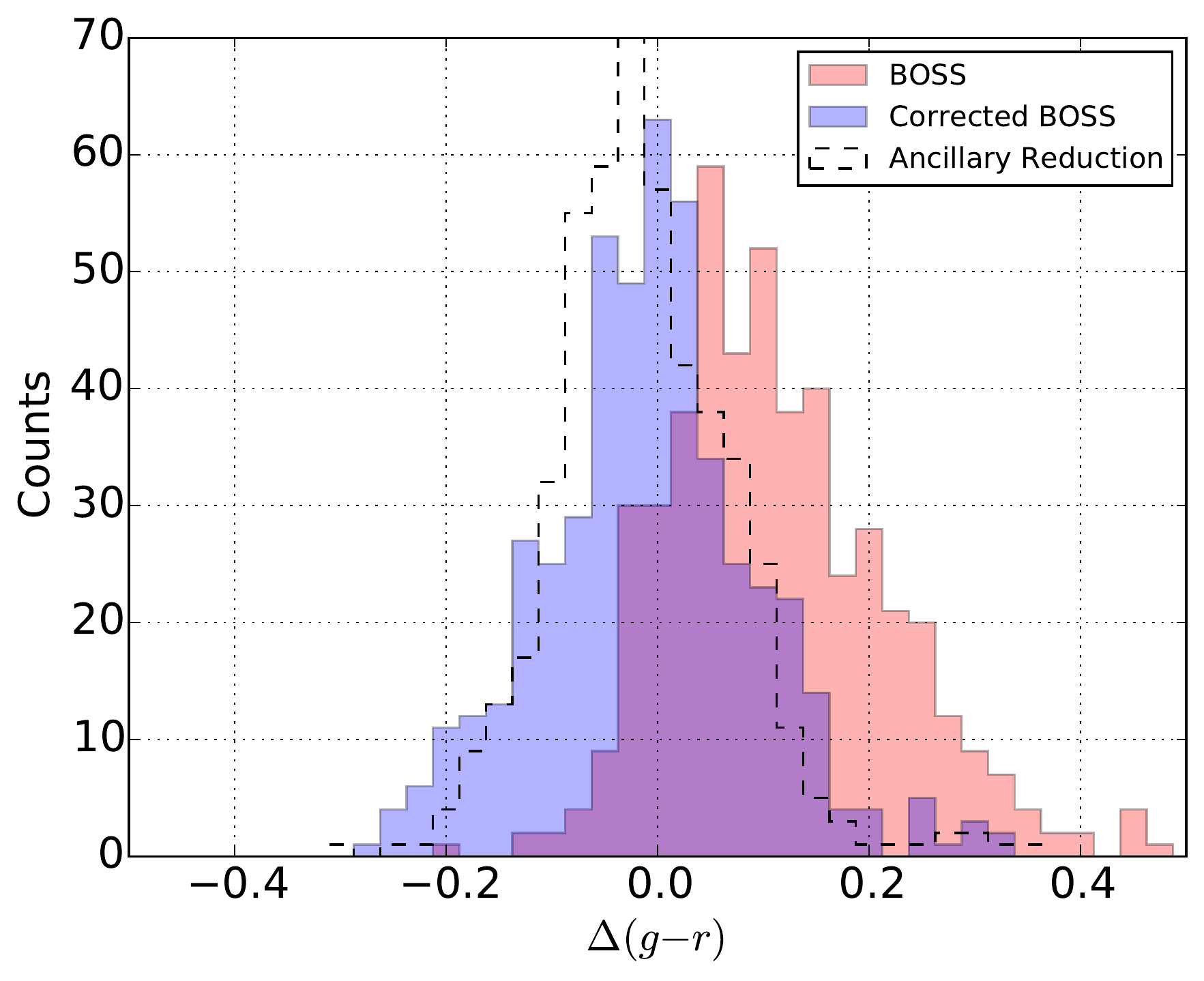}
\caption{Residual $(g-r)$ color distribution between SDSS imaging and BOSS synthetic magnitudes for a sample of offset standard stars. The red and blue distributions correspond to synthetic magnitudes calculated from the BOSS DR12 spectra before and after the correction, respectively. The black dashed line corresponds to synthetic magnitudes calculated from the modified data processing pipeline that uses offset standard stars for spectrophotometric calibration. \label{fig:mag-compare} }
\end{figure}

Following the assessment of spectrophotometric calibration performed in \citet{2013AJ....145...10D}, we calculate synthetic AB magnitudes \citep{1983ApJ...266..713O} using SDSS filter-curve-weighted\footnote{\url{http://www.sdss.org/instruments/camera/\#Filters}} integration of spectra for BOSS targets and compare them to their targets' corresponding SDSS imaging magnitudes \citep{1998AJ....116.3040G, 1996AJ....111.1748F}. We apply AB magnitude corrections for SDSS imaging magnitudes ($m_{AB} - m_{SDSS}$) used by the BOSS data processing pipeline for the $g$-, $r$-, and $i$-band of 0.036, 0.015, and 0.013. 
The BOSS spectrophotometric calibration of standard stars yields $(g - r)$ colors that are on average $0.022 \pm 0.003$ magnitudes redder compared to their SDSS imaging magnitudes. Note that we calculate the sample dispersion using the normalized median absolute deviation.
For the offset standards, the $(g - r)$ colors are $0.117 \pm 0.004$ magnitudes bluer before applying the correction and $0.006 \pm 0.004$ magnitudes redder after. Using spectra from the validation pipeline yields $(g - r)$ colors that are $0.010 \pm 0.003$ magnitudes redder. 
\fig{mag-compare} displays the distribution of color residuals $\Delta(g - r)$ between SDSS imaging and BOSS synthetic magnitudes for the offset standards. 
The red and blue distributions correspond to synthetic magnitudes calculated from the BOSS DR12 spectra before and after applying the corrections, respectively. 
The black dashed line corresponds to synthetic magnitudes calculated using spectra from the validation pipeline that uses offset standard stars for spectrophotometric calibration. 


A complete listing of magnitude and color residuals for the relevant data samples is given in \tab{color-compare}. 
Our corrections generally remove a significant bias in the mean and reduce the dispersion of the magnitude and color residuals. 

\begin{table*}
\small
\centering
\begin{tabular}{ l r  r r  r r  r r  r r  r r }
    \hline
    Data Sample & \# Targets & \multicolumn{2}{ c }{$\Delta(g-r)$} & \multicolumn{2}{ c }{$\Delta(r-i)$} & \multicolumn{2}{ c }{$\Delta g$} & \multicolumn{2}{ c }{$\Delta r$} & \multicolumn{2}{ c}{$\Delta i$} \\
    & & \multicolumn{1}{c}{Mean} & \multicolumn{1}{c}{Disp.} & \multicolumn{1}{c}{Mean} & \multicolumn{1}{c}{Disp.} & \multicolumn{1}{c}{Mean} & \multicolumn{1}{c}{Disp.} & \multicolumn{1}{c}{Mean} & \multicolumn{1}{c}{Disp.} & \multicolumn{1}{c}{Mean} & \multicolumn{1}{c}{Disp.} \\ 
    \hline
    Spec. standards        & 400  & $ 0.022$ & $ 0.057$ & $ 0.014$ & $ 0.032$ & $ 0.032$ & $ 0.057$ & $ 0.011$ & $ 0.052$ & $-0.004$ & $ 0.060$ \\ 
    Offset standards       & 486  & $-0.117$ & $ 0.098$ & $-0.053$ & $ 0.068$ & $ 0.084$ & $ 0.122$ & $ 0.201$ & $ 0.159$ & $ 0.255$ & $ 0.214$ \\ 
    Corr. offset standards & 486  & $ 0.006$ & $ 0.085$ & $ 0.004$ & $ 0.062$ & $ 0.047$ & $ 0.124$ & $ 0.041$ & $ 0.163$ & $ 0.037$ & $ 0.198$ \\ 
    Spec. offset standards & 486  & $ 0.010$ & $ 0.077$ & $ 0.008$ & $ 0.058$ & $ 0.054$ & $ 0.086$ & $ 0.044$ & $ 0.097$ & $ 0.036$ & $ 0.127$ \\ 
    Failed quasars         & 1,049 & $-0.067$ & $ 0.132$ & $-0.065$ & $ 0.127$ & $ 0.071$ & $ 0.158$ & $ 0.138$ & $ 0.187$ & $ 0.203$ & $ 0.245$ \\ 
    Corr failed quasars    & 1,049 & $ 0.058$ & $ 0.135$ & $-0.006$ & $ 0.116$ & $ 0.029$ & $ 0.157$ & $-0.028$ & $ 0.186$ & $-0.023$ & $ 0.238$ \\
    \hline
\end{tabular}
\caption{Band-pass filter magnitude and color differences between SDSS imaging and synthetic AB magnitudes calculated from BOSS spectra. The sample mean and dispersion are listed for various data samples. Dispersions are calculated as the normalized median absolute deviation. \label{tab:color-compare}}
\end{table*}

\section{Discussion} \label{sec:discussion}

The corrections described here provide a significant improvement over the uncorrected quasar spectra in DR12 and are publicly available\footnote{\url{http://darkmatter.ps.uci.edu/tpcorr/}.}. However, there are limitations to attempting to correct the quasar spectrophotometric calibration as a post-processing step after the data processing pipeline, since many simplifying assumptions are required. 
In order to achieve better results, it is necessary and desirable to use as much of the existing pipeline calibration algorithms as possible but with spectrophotometric standards observed with the same focal-plane offsets as the quasar targets. 
Short of including spectrophotometric standards with focal plane offsets, the throughput corrections described here can be applied as a preprocessing step of the calibration process. 
Other improvements over the method described here include using a more realistic description for the PSF, using the actual guiding adjustments for determining centroid offsets, and applying corrections to individual exposures prior to co-addition. 
This is especially relevant for the extended Baryon Oscillation Spectroscopic Survey (eBOSS) \footnote{\url{http://www.sdss.org/surveys/eboss/}} of SDSS-IV, which uses similar focal plane offsets for quasar targets. 
We are currently working with the eBOSS data processing team to implement the throughput corrections described for individual exposures into the data processing pipeline. 
The proposed Dark Energy Spectroscopic Instrument \citep[DESI;][]{2013arXiv1308.0847L} includes an atmospheric dispersion compensator in its design so the effects of ADR should be much reduced for that survey.

\acknowledgments

Funding for SDSS-III has been provided by the Alfred P. Sloan Foundation, the Participating Institutions, the National Science Foundation, and the U.S. Department of Energy Office of Science. The SDSS-III web site is \url{http://www.sdss3.org/}.

SDSS-III is managed by the Astrophysical Research Consortium for the Participating Institutions of the SDSS-III Collaboration including the University of Arizona, the Brazilian Participation Group, Brookhaven National Laboratory, Carnegie Mellon University, University of Florida, the French Participation Group, the German Participation Group, Harvard University, the Instituto de Astrofisica de Canarias, the Michigan State/Notre Dame/JINA Participation Group, Johns Hopkins University, Lawrence Berkeley National Laboratory, Max Planck Institute for Astrophysics, Max Planck Institute for Extraterrestrial Physics, New Mexico State University, New York University, Ohio State University, Pennsylvania State University, University of Portsmouth, Princeton University, the Spanish Participation Group, University of Tokyo, University of Utah, Vanderbilt University, University of Virginia, University of Washington, and Yale University.

\bibliographystyle{apj}
\bibliography{arxiv}

\end{document}